# ARTICLE

# A phase-field model for the evaporation of thin film mixtures

Olivier J.J. Ronsin[a]*, DongJu Jang[b], Hans-Joachim Egelhaaf[b], Christoph J. Brabec[c,d] and Jens Harting[a,e]*

The performance of solution-processed solar cells strongly depends on the geometrical structure and roughness of the photovoltaic layers formed during film drying. During the drying process, the interplay of crystallization and liquid-liquid demixing leads to the structure formation on the nano- and microscale and to the final rough film. In order to better understand how the film structure can be improved by process engineering, we aim at theoretically investigating these systems by means of phase-field simulations. We introduce an evaporation model based on the Cahn-Hilliard equation for the evolution of the fluid concentrations coupled to the Allen-Cahn equation for the liquid-vapour phase transformation. We demonstrate its ability to match the experimentally measured drying kinetics and study the impact of the parameters of our model. Furthermore, the evaporation of solvent blends and solvent-vapour annealing are investigated. The dry film roughness emerges naturally from our set of equations, as illustrated through preliminary simulations of spinodal decomposition and film drying on structured substrates.

## Introduction

In many practical applications, thin films can be produced by solution processing: the materials are dissolved in a solvent or solvent blend and deposited on the desired substrate. The solvents then evaporate, leading to the formation of the dry functional film. These solvent-based production methods have applications in the fields of coatings, membrane fabrication for energy technologies such as batteries [1] or fuel cells [2] [3], or deposition of photoactive layers for solution processed organic electronics [4] and photovoltaics [5] [6]. They allow for highly efficient, eco-friendly and cheap fabrication processes: the solution can be deposited by various techniques such as spin-coating, doctor blading, slot-die coating or inkjet printing and dried at relatively low temperatures, typically below 150°C. Most of these fabrication techniques are compatible with industrial roll-to-roll processes and enable a straightforward upscaling of the production [7] [8] [9]. The material systems of interest might be quite different in these various applications, but complex physical processes often occur during drying, leading to the self-organization of the functional layer. Examples of these processes are the self-assembly and interactions of nanoparticles, for instance during Li-ion battery electrode production [10] [11] or colloidal particle coatings [12], spinodal decomposition of polymer blends for organic electronics [13] [14] [15] [16], self-structuring of block-copolymers, and polycrystalline nucleation and growth of photoactive layers in perovskite solar cells [17] [18].

The properties of the final film usually strongly depend on its structure. In the field of solution processed photovoltaics, which is the main concern of this work, the process-structure-property relationship is a key factor for improving the efficiency of the solar cells. In the case of organic photovoltaics (OPV), high-performance solar cells are based on the so-called bulk-heterojunction (BHJ) structure, where the organic donor and acceptor are separated at the nanometer scale. For maximum device performance, an optimal BHJ morphology is required, which on the one hand provides sufficient proximity of donor and acceptor components to allow for efficient exciton dissociation at the donor/acceptor interface, and on the other hand also exhibits sufficient phase separation to form bi-continuous percolation pathways for the extraction of the charges formed. Thereby it is important to produce crystalline or ordered domains in order to obtain sufficient mobilities for the charge carriers [19]. These sophisticated structures form spontaneously during the drying of the wet-deposited solution and strongly influence the solar cell efficiency. In the case of perovskite photoactive layers, perovskite crystals grow in the evaporating solution and recent literature indicates that there is a close correlation between the crystallization process upon solution deposition, the resulting perovskite crystal structure, the defect density and the efficiency of the completed solar cell [20] [21] [22]. It has been additionally argued by many authors, referring to the general knowledge of crystallization processes, that the crystal structure depends on the relative speed of nucleation, crystal growth, solute diffusion and solvent evaporation processes [23] [24] [25]. As a consequence, meticulous processing strategies, such as e.g. thermal or solvent annealing treatment, binary or ternary solvent formulation, and thermally assisted film drying are currently being developed [26] [27] [28] [29] [30].

[a.] Helmholtz Institute Erlangen-Nürnberg for Renewable Energy, Forschungszentrum Jülich, Fürther Straße 248, 90429 Nürnberg, Germany
[b.] ZAE Bayern—Solar Factory of the Future, Energy Campus Nürnberg, Fürther Straße 250, 90429 Nürnberg, Germany
[c.] Institute of Materials for Electronics and Energy Technology (i-MEET), Friedrich-Alexander-Universität Erlangen-Nürnberg, 91058 Erlangen, Germany
[d] Helmholtz Institute Erlangen-Nürnberg for Renewable Energy, Forschungszentrum Jülich, Immerwahrstrasse 2, 91058 Erlangen, Germany
[e] Department of Applied Physics, Eindhoven University of Technology, PO box 513, 5600MB Eindhoven, The Netherlands

In order to reach significant performance improvements, it is essential to gain full control of the impact of the processing conditions on the formation of the optimum BHJ and perovskite morphologies. In particular, a better understanding of the physics driving the active layer formation is strongly needed to devise a rationale for the development of materials and processes for organic and perovskite solar cells. This is a challenge, due to the complexity of the drying process, which includes physical transformations such as crystallization and/or liquid-liquid and liquid-solid phase separation. Moreover, the final system remains quenched far from equilibrium and the resulting morphology is kinetically controlled [31]. Thus, in order to understand the complex physics behind morphology formation, not only the thermodynamics but also the rate constants of each transformation have to be determined. This renders simulations of great interest for the fundamental understanding of the formation of active layers.

In the last decade, different simulation approaches have been proposed to contribute to the understanding of the active layer formation in photovoltaic systems. At small scales, coarse-grained molecular dynamics (CGMD) [32,33,34], dissipative particle dynamics (DPD) [35,36,37] or self-consistent field theory (SCFT) [38,39] studies have been performed. However, the length scales and predominately timescales accessible with such techniques prevent them from being used for a kinetic description of the whole drying process, though simulations of about $(100nm)^3$ volumes may be performed. To deal with kinetic aspects and the explicit description of solvent evaporation, the system has to be described at a larger scale within the framework of continuum mechanics. Although the lattice Boltzmann method (LBM) could be a method of choice for investigating phase separating fluid mixtures [40,41,42], crystallization [43,44,45] and evaporation [46,47,48], almost all simulation studies performed so far in the field of optoelectronics are based on phase-field modelling. As the LBM, the phase-field method is a well-established technique for solving interfacial problems with diffuse interfaces. It has been widely applied to various problems, including crystallization in alloys or liquid-liquid phase separation [49,50,51]. The use of phase-field modelling of drying films and more specifically in the field of organic electronics remains however limited to a few papers exclusively dealing with spinodal decomposition of liquid mixtures. Wodo et al. dealt extensively with ternary systems including a polymer, a fullerene and an evaporating solvent, adding also specific interactions with the substrate [52,53]. They gained insight into the impact of process parameters on the final amorphous two-phase structure. Michels et al. investigated donor/acceptor mixtures with an evaporating solvent [13,54,55,56]. Aside from instructive physical analysis of the evaporation-induced spinodal decomposition, they proposed first qualitative comparisons with experimental data on real systems. However, even if many organic optoelectronic active materials are at least semi-crystalline, crystallization processes have not been considered.

In numerous simulation studies of drying thin films that rely on the assumptions of continuum mechanics, either related to diffusion phenomena in miscible liquids [57], in phase-field simulations for amorphous incompatible systems [13,53,54,55,56,58] or crystalline systems [59,60,61], the simulation box is restricted to the liquid phase. Evaporation is taken into account as a boundary condition at the top of the film: the solvent flow going out of the system is assumed to be proportional to the difference between the solvent concentration at the film surface and its vapor pressure. This very simple and efficient procedure allows to gain deep insight into the microstructure evolution. However, the roughness of the final dry solid film cannot be investigated. Sophisticated models have been developed to describe the physics of evaporation. Borcia and Besterhorn proposed a single component phase-field model with the fluid density as non-conserved order parameter to write the free energy of the system. The resulting non-classical phase-field term is then being added to the Navier-Stokes equations. They investigated bubble generation in liquids, droplets on substrates and Marangoni convection with their model [62,63,64]. Badillo [65] and Kaempfer and Plapp [66] also proposed diffuse interface models of liquid-vapor phase-transition with a phase field approach coupled to the Navier-Stokes equations and taking into account temperature changes. Recently, Cummings et al. proposed a similar approach taking into account solvent density variations and hydrodynamic effects for three-component systems, namely one solvent and two immiscible materials, and demonstrated simulations of spinodal decomposition, including the appearance of rough final films [67]. These sophisticated models focus on the proper description of the phase change but might not be well suited for the investigation of self-structuring films: the remarkable effort for describing properly the liquid-vapor interface is annihilated by the fact that it is in practice numerically almost intractable to use the correct diffusion coefficients in the liquid and gas phase at the same time (with four orders of magnitude difference). As a consequence, simulations with sufficiently small time steps to describe diffusion in the gas phase and with sufficient total time to describe the microstructure evolution are hardly accessible. Finally, Schweigler et al. [68] proposed a simple evaporation model for droplets on complex substrates. The phase of the liquid is described with the help of a non-conserved order parameter whose kinetics follows the Allen-Cahn equation. The comparison of their simulations with experimental data is very promising.

Our goal is to investigate the structure formation of perovskite and semi-crystalline OPV photoactive layers upon drying. This implies taking into account liquid-liquid phase separation and growth, as well as polycrystalline nucleation and growth, in multicomponent polymer mixtures, potentially with polymeric materials. These physical transformations are initiated by the solvent removal and a proper description of the evaporation process is therefore crucial to properly describe these kinetically driven systems, as well as to define optimal processing strategies. In this paper, we will only focus on the description of a simple phase-field model for the evaporation of multicomponent mixtures.

It should be emphasized again that, even if we deal here with an evaporation model, the structure formation in the liquid film remains our only concern. Now, the evaporation mainly has an impact on the structure formation through its drying rate. The gas fluxes in the low-density vapor phase itself are not expected to strongly influence the evolution of the liquid phase unless they induce a change in the drying rate. The liquid-vapor interface can be considered as a boundary condition for the structuring thin-film and its detailed structure should not play a major role in the microstructure evolution. As a consequence, we neither aim at nor pretend describing properly the thermodynamics of the liquid-vapor phase transition and the diffusion processes in the gas phase. In contrast to that, we demand the model to have the following properties:

It shall be able to handle multicomponent mixtures, and in particular solvent blends because this is widely used in the field of solution-processed photovoltaics to improve the final microstructures. Moreover, even if this will not be discussed further in this paper, it should be able to describe properly the interactions of the solvents with the other species.

Investigation of the roughness of the dry film should be made possible. This is fundamental especially for crystallizing systems such as photoactive perovskite films: depending on the processing parameters, the roughness can reach several tenth of nanometers, or the substrate might even not be fully covered, leading to dramatic variations in the photovoltaic performance.

The drying speed and total drying time are of primary importance for the generation of the final structure. In systems where liquid-liquid spinodal decomposition takes place, domains of the separated phases can grow until they are kinetically quenched because of the viscosity increase due to the decrease of solvent concentration. In crystallizing systems, as originally proposed by Lamer and Dinegar [69], the evolution of the solute concentration depends on the balance between the solvent evaporation, which tends to increase the solute volume fraction, and the crystallization process which consumes solute material from the liquid phase. How long the concentration stays above the nucleation and growth threshold depends on the relative evaporation, nucleation and growth rates [24]. The validity of the simulated evaporation kinetics, and its correct sensitivity to material parameters is thus very important.

In the proposed model, we neither take into account the density variation upon phase change nor the hydrodynamics in the gas phase, so that on the one hand, the proposed model is quite simple as compared to the approaches proposed by Cummings et al. [67], Borchia and Besterhorn [62,63,64] or Kaempfer and Plapp [66]. On the other hand, compared to the classical approach with the gas flow as a boundary condition at the top of the liquid film, the film roughness will be naturally obtained, and we will show that the evaporation kinetics matches experimental data much better.

## The phase-field model

### The free energy functional

The kinetic evolution of the system towards its thermodynamic equilibrium is simulated within a phase-field framework. In this approach, we describe the system with the help of its free energy. Without loss of generality, the free energy functional reads

$$G_{tot} = \int_V \left( \Delta G_V^{loc} + \Delta G_V^{nonloc} \right) dV , \quad (1)$$

where $V$ is the system volume. $\Delta G_V^{loc}$ is the local free energy density and $\Delta G_V^{nonloc}$ the non-local contribution due to the field gradients. The system is composed of $n-1$ solutes and solvents plus one additional fluid which describes the ambient air, i.e. a total number of $n$ fluids. The free energy depends not only on the respective volume fractions $\varphi_i$ of all fluids in the system, but also on the order parameter $\Phi_{vap}$ whose value varies between 0 in the liquid phase and 1 in the gas phase. $\Phi_{vap}$ describes the phase for all the $n_{solv}$ volatile species at the same time. An additional order parameter $\Phi_{air}$ is introduced for the description of the air. The local part of the free energy is given by

$$\Delta G_V^{loc}(\{\varphi_i\}, \Phi_{vap}) = \Delta G_V^{ideal}(\{\varphi_i\}) + \Delta G_V^{inter}(\{\varphi_i\}, \Phi_{vap}) + \Delta G_V^{LV}(\{\varphi_i\}, \Phi_{vap}). \quad (2)$$

The first term on the right-hand side of the equation above represents the free energy density change upon ideal mixing,

$$\Delta G_V^{ideal}(\{\varphi_i\}) = \frac{RT}{v_0} \sum_{i=1}^n \frac{\varphi_i \ln \varphi_i}{N_i}, \quad (3)$$

with $R$ being the gas constant and $T$ the temperature. $v_0$ is the molar volume of the lattice site considered to calculate the free energy of mixing in the sense of the Flory-Huggins theory [70] and $N_i$ is the molar size of the fluid $i$ in terms of units of the lattice site volume, so that its molar volume is $v_i = N_i v_0$. The second term stands for the interactions between the fluids:

$$\Delta G_V^{inter}(\{\varphi_i\}, \Phi_{vap}) = \frac{RT}{v_0} \begin{pmatrix} \sum_{i=1}^n \sum_{j>i}^n \varphi_i \varphi_j \chi_{ij,ll} \\ + \Phi_{vap}^2 \sum_{k \in \{solv\}} \sum_{\substack{j \neq k}}^n \varphi_k \varphi_j \chi_{kj,vl} \\ + \Phi_{vap}^2 \sum_{k \in \{solv\}} \sum_{\substack{j \in \{solv\} \\ j \neq k}} \varphi_k \varphi_j \chi_{kj,vv} \\ + \Phi_{air}^2 \varphi_{air} \sum_{\substack{j \neq k}}^n \varphi_j \chi_{airj,vl} \\ + \Phi_{air} \Phi_{vap} \varphi_{air} \sum_{j \in \{solv\}} \varphi_j \chi_{airj,vv} \end{pmatrix} \quad (4)$$

This form is inspired by a generalization of the classical Flory-Huggins theory originally proposed by Matkar and Kyu for binary systems with crystalline materials [71,72]. The first term is the standard Flory-Huggins interaction term with $\chi_{ij,ll}$ being the interaction parameter between the liquid phases of fluids $i$ and $j$. The solvents might have a vapor phase and these vapor phases might interact with liquid phases, especially in the diffuse liquid-vapor interface. The second term stands for these interactions, with $\chi_{kj,vl}$ being the interaction parameter between the vapour phase of fluid $k$ and the liquid phase of fluid $j$. This term can be understood considering that $\Phi_{vap}$ can be interpreted as the proportion of solvent being already in the vapor phase, so that $\varphi_k \Phi_{vap}$ is the quantity of vapour and $\varphi_j \Phi_{vap}$ the amount of liquid phase interacting with this vapour [72]. The third term stands for the vapor-vapor interactions, with $\chi_{kj,vv}$ being the interaction parameter between the vapor phases of the fluids $k$ and $j$. The two last terms are similar to the second and third one but correspond to the contribution of the air. Following Matkar and Kyu, we use $\chi_{kj,vv} = c\sqrt{\chi_{kj,lv}}\sqrt{\chi_{jk,lv}}$ with $c$ ranging from $-2$ for fully compatible vapors to $0$ for fully incompatible vapors.

The last term on the RHS of Eq. *(2)* represents the free energy density of phase change, similar to what is often used for the simulation of crystallization [49] [50],

$$\Delta G_V^{LV}(\{\varphi_i\}, \Phi_{vap}) = \sum_{k \in \{solv\}} \rho_k \varphi_k (g(\Phi_{vap})H_k + p(\Phi_{vap})F_k) \quad (5)$$

Here, $\rho_k$ is the density of the fluid $k$. $F_k$, the free energy density generating the driving force for evaporation of the fluid $k$, will be abusively called "driving force for evaporation" in the following for simplicity and will be discussed later in more detail. The parameter $H_k$ generates an energy barrier for the liquid-vapour phase transition provided that $|3F_k/H_k| < 1$. $p(\Phi_{vap})$ and $g(\Phi_{vap})$ are the following interpolation functions (note that other choices are possible without significant impact on the model behavior):

$$\begin{cases} g(\Phi_{vap}) = \Phi_{vap}^2 (\Phi_{vap} - 1)^2 \\ p(\Phi_{vap}) = \Phi_{vap}^2 (3 - 2\Phi_{vap}) \end{cases} \quad (6)$$

Finally, the non-local part of the free energy functional describes the surface tension contributions due to the concentration gradients and due to the order parameter gradient as

$$\Delta G_V^{nonloc}(\{\nabla \varphi_i\}, \nabla \Phi_{vap}) = \frac{1}{2} \sum_{i=1}^{n} \kappa_i (\nabla \varphi_i)^2 + \frac{1}{2} \varepsilon_{vap}^2 (\nabla \Phi_{vap})^2, \quad (7)$$

where $\kappa_i$ is the surface tension parameter for the concentration gradient of fluid $i$ and $\varepsilon_{vap}$ the surface tension parameter for the gradient of the order parameter. The surface tension between two liquid phases $i$ and $j$ is proportional to $\sqrt{\kappa_i + \kappa_j}$, whereas the surface tension of a liquid-vapor interface also contains a contribution from the phase variation. However, the surface tension also depends on the other thermodynamic properties such as the molar volumes and interaction parameters and can be computed with standard methods described elsewhere [50] [51].

**Kinetic equations**

The kinetic equations for the volume fractions, which are conserved quantities in our model, are based on the formalism initially proposed by Cahn and Hilliard for binary mixtures [73] [74] and generalized later for multicomponent mixtures [54] [56] [75] [76] [77] [78] [79]:

$$\frac{\partial \varphi_i}{\partial t} = \frac{v_0}{RT} \nabla \left[ \sum_{j=1}^{n-1} \Lambda_{ij} \nabla \left( \mu_{V,j}^{gen} - \mu_{V,n}^{gen} \right) \right] \quad i = 1 \dots n-1 \quad (8)$$

This can be understood as a set of continuity equations, where the fluid fluxes are proportional to the gradient of the exchange chemical potential which is the driving force for the system evolution. In the form of the equation above, $\mu_{V,j}^{gen}$ is a chemical potential density, basically the derivative of the free energy density, but it is generalized to take into account the non-local part of the free energy functional in the following way:

$$\mu_{V,j}^{gen} - \mu_{V,n}^{gen} = \left(\frac{\partial \Delta G_V}{\partial \varphi_j}\right) - \left(\frac{\partial \Delta G_V}{\partial \varphi_n}\right) \\ - \left(\nabla \left(\frac{\partial \Delta G_V}{\partial (\nabla \varphi_j)}\right) - \nabla \left(\frac{\partial \Delta G_V}{\partial (\nabla \varphi_n)}\right)\right) \quad (9)$$

The first two terms correspond to the "usual" chemical potential, whereas the two last contributions take into account the potential due to concentration gradients. The $\Lambda_{ij}$ are mobility coefficients related to the diffusion coefficients in a non-trivial way. First, they have to ensure the Onsager relationship, $\Lambda_{ij} = \Lambda_{ji}$. Second, in order to ensure the Gibbs-Duhem relationship together with the incompressibility constraint (recall that we assume here constant densities also in the vapor phase), it can be shown that the fluxes are fully coupled and that the mobility coefficients are volume fraction dependent. Several theories have been proposed to derive correct expressions for the flux. The most successful ones are the so-called "slow mode theory", initially proposed by De Gennes [80] and the "fast-mode theory" proposed by Kramer et al. [81]. They lead to different expressions of the mobility coefficients and it turns out that for the "slow-mode theory", the mutual diffusion coefficient in a binary system is controlled by the slowest component, while it is controlled by the fastest component in the "fast-mode theory". These considerations gave the names for both theories. Even if the controversy between both theories is not fully resolved, the fast-mode theory seems to better match experimental data and can also be derived from the general Maxwell-Stefan equations framework[82], so that we choose to use the fast-mode theory which gives

$$\begin{cases} \Lambda_{ii} = (1-\varphi_i)^2 M_i + \varphi_i^2 \sum_{k=1, k \neq i}^{n} M_k \\ \Lambda_{ij} = -(1-\varphi_i)\varphi_j M_i - (1-\varphi_j)\varphi_i M_j + \varphi_i \varphi_j \sum_{\substack{k=1, \\ k \neq i \neq j}}^{n} M_k \end{cases} \quad (10)$$

with the coefficients $M_i$ related to the fluid self-diffusion coefficients $D_{s,i}$ through

$$M_i = \frac{v_0}{RT} N_i \varphi_i D_{s,i}(\varphi_i). \quad (11)$$

In general, the self-diffusion coefficients themselves also depend on the mixture composition, but they will be assumed to be constant in this paper for simplicity. The order parameter follows the Allen-Cahn equation,

$$\frac{\partial \Phi_{vap}}{\partial t} = -M\left(\frac{\partial \Delta G_V}{\partial \Phi_{vap}} - \nabla\left(\frac{\partial \Delta G_V}{\partial(\nabla \Phi_{vap})}\right)\right), \quad (12)$$

whereby $M$ is the mobility coefficient for the liquid-vapor interface that fixes the evaporation rate. There is no kinetic equation for the air order parameter, meaning that we do not minimize the free energy relative to $\Phi_{air}$. Instead, we simply assume that $\Phi_{air} = \Phi_{vap}$, so that there is only a single vapor phase in the system.

Both Cahn-Hilliard and Allen-Cahn equations ensure that the system progressively minimizes its free energy relative to the volume fractions and the order parameter, i.e. it relaxes towards its thermodynamic equilibrium.

**The evaporation model**

The driving force for evaporation $F_k$ deserves some additional comments: the driving force for phase change is supposed to be the free energy difference between both phases. Thus, from the known free energy of mixing in the liquid and gas phases we can write

$$F_k \sim ln\left(\frac{\varphi_k}{\varphi_{sat,k}}\right) - \frac{ln\varphi_k}{N_k}, \quad (13)$$

with $\varphi_{sat,k}$ being the partial vapour pressure of the solvent $k$. The first term here is the free energy of mixing in the gas phase and the second term cancels the ideal mixing term coming from Eq. *(3)*. With such a driving force, it can be shown that liquid-vapor equilibrium properties such as Raoult's law for ideal gas mixtures and various Henry's laws for non-ideal mixtures (depending on the interaction parameters and molar volumes of solvents) can be recovered. However, the speed of an interface whose kinetics is described by the Allen-Cahn equation is known to vary linearly with $F_k$ [49,50,83] and this still holds if the Allen-Cahn equation is coupled to the Cahn-Hilliard equation for the volume fractions. This means that the evaporation rate varies linearly with $ln(\varphi_{sat,k})$ which is in contradiction with the experimental results clearly indicating a linear relationship between the evaporation rate and the vapor pressure [84]. Actually, evaporation is assumed to be a two-step process: the liquid-vapor phase transition occurs at the surface of the liquid film, and a vapor layer in equilibrium with the liquid forms above the liquid film, hence with a partial pressure equal to the solvent vapor pressure $P_{sat}$. This gas layer then exchanges molecules with the environment where the solvent pressure is different (noted $P_{vap}$). It turns out that the second step is the limiting one and determines the evaporation rate, which is in fact kinetically driven rather than thermodynamically. The exchange process between the gas layer and the environment has been originally derived by Hertz and Knudsen and has been improved over time [85]. In the current work, we write the driving force according to the Hertz-Knudsen formula as

$$F_k = -a_k\sqrt{\frac{N_k v_0}{RT}}(\varphi_{sat,k} - \varphi_{vap,k}), \quad (14)$$

where $\varphi_{vap,k}$ is the partial pressure of the solvent $k$ in the environment and $a_k$ is a kinetic model parameter. This ensures that the evaporation rate varies linearly with the pressure difference.

As in the classical Allen-Cahn model, this term is sufficient to drive the liquid-vapor phase transition as soon as it is negative. But in a closed system, this results in an equilibrium between the liquid and the gas phase, so that the solvent cannot be fully removed from the film, even if $\varphi_{vap} < \varphi_{sat}$, which is unphysical. Therefore, we propose to add a flow boundary condition driving the solvents out of the system at the top of the simulation box. Since the volume fractions are conserved quantities and the sum of all volume fractions has to be equal to one, this has to be compensated by an inflow for another fluid, namely the ambient air. The boundary condition for the solvent is written as

$$j_k = b_k \frac{h_{v,0}}{h_v(t)}(\varphi_{top,k} - \varphi_{vap,k}). \quad (15)$$

Here, $\varphi_{top,k}$ is the solvent concentration at the top of the simulation box, $b_k$ a coefficient to adjust the flow rate, $h_{v,0}$ the initial vapor film thickness and $h_v(t)$ the mean vapor film thickness at time $t$. Since we simulate the drying of a liquid film located in an infinitely large vapor phase, we wish the boundary condition to induce a constant solvent flow at the film surface. $h_{v,0}/\overline{h_v(t)}$ is a correction factor in order to ensure this condition. Otherwise, the mean concentration gradient in the gas phase between the liquid film surface and the simulation box boundary, and hence the solvent flux at the liquid film surface, decreases inversely proportional to $h_v(t)$, an effect that has been outlined in [46].

## Model behavior

The equations *(1)-(12)* and *(14)-(15)* are written in a dimensionless form using $\widetilde{\Delta G_V^{loc}} = \Delta G_V^{loc}/g_{sc}$, $\widetilde{\Delta G_V^{nonloc}} = \Delta G_V^{nonloc}/(g_{sc}l_{sc}^2)$, $\tilde{l} = l/l_{sc}$, $\widetilde{\Lambda_{ij}} = \Lambda_{ij}/D_{sc}$ and $\tilde{t} = t/t_{sc}$, The coefficients $g_{sc}$, $l_{sc}$, $D_{sc}$ and $t_{sc}$ are chosen as $g_{sc} = RT/v_0$, $l_{sc} = \sqrt{max(\kappa_{1...n}, \varepsilon_{vap})/g_{sc}}$ to be consistent with the size of the thinnest interface of the system, $D_{sc} = max(N_i D_{s,i})$ and finally $t_{sc} = l_{sc}^2/D_{sc}$. The equations are numerically solved using an Euler explicit finite difference scheme with variable time steps chosen as follows: in order to ensure numerical stability, the time step has to be smaller than $dx^4/(2^{2d+1}max(\Lambda_{ij})max(\kappa_i + \kappa_j))$ for the Cahn-Hilliard equation, and smaller than $dx^2/(2^d M \varepsilon_{vap}^2)$ for the Allen-Cahn equation at any time. Since the mobilities are composition-dependent, both values may vary with time. They are calculated at each time step and we simply fix the time step to be 90% of the smallest one. Note that the dynamics and maximum time step related to both equations might differ by orders of magnitude, with the one for the Cahn-Hilliard equation being usually the limiting one. We perform one-dimensional simulations with a grid of 200 mesh points and a grid resolution of 3nm with one solute and one solvent. We deliberately choose to simulate simplified model systems in order to focus on the basic physics and model behavior. As a consequence, the parameters used in this work are not expected to be representative of a particular real system and might differ from experimental values. Real systems are much more complex, with different parameters for different materials, and the evaluation of the input parameters is far from trivial (interaction parameters, composition-dependent diffusion coefficients, surface tension parameters) and requires additional work. Performing quantitative simulations with realistic input parameters is beyond the scope of this paper and will be the topic of future work. Here, all molar volumes are equal, the liquid phases are perfectly miscible, and the solvent and the air are made perfectly miscible in the gas phase by choosing $\chi_{kj,vv} = -2\chi_{kj,lv} = -2\chi_{kj,vl}$ for all vaporizable species $k$ and $j$. The vapor and liquid phase are strongly immiscible. The surface tension parameters and diffusion coefficients are also the same for all fluids. The remaining parameters are varied and it will be specified when they differ from the basic parameter set summarized in Table 1. The grid spacing is adapted to the liquid-vapor interface thickness and is set to 3nm.

| T | 300 K | $F_{solv}$ | -90 kJ/kg |
|---|---|---|---|
| $v_0$ | $10^{-4}$ m³/mol | $F_{air}$ | -300 kJ/kg |
| $N_i$ (all fluids) | 1 | $H_{solv}$ | 500 kJ/kg |
| $\rho_i$ (all fluids) | 1000 kg/m³ | $H_{air}$ | 500 kJ/kg |
| $\chi_{ii,ll}$ (all) | 0 | $\varepsilon_{vap}$ | $6.5.10^{-5}$ (J/m)$^{0.5}$ |
| $\chi_{kj,vl}$ (all) | 5 | M | $4.5.10^4$ s$^{-1}$ |
| $\chi_{kj,vv}$ (all) | -10 | $\varphi_{vap}$ | 0 |
| $\kappa_i$ (all) | $2.10^{-9}$ J/m | b | 3 J.m$^{-4}$ |
| $D_{s,i}^{\varphi_k \to 1}$ (all) | $10^{-8}$ m²/s | | |

Table 1: basic parameter set used in the 1D simulations for the investigation of the model behavior

Figure 1 shows the typical evolution of all the fields for a one-dimensional simulation of a system with one solute and one solvent. The initial solvent concentration in the film is 90% and the interface moves from the right to the left with time. The evaporation rate in this simulation is sufficiently low so that the volume fractions are almost homogeneous in the liquid film. The solute concentration increases while the solvent concentration decreases. The global solvent quantity in the system also decreases thanks to the flux boundary condition, compensated by the air inflow. Almost no solvent remains either in the dry film or in the vapor at the end (note that the simulation has been interrupted before the final state is reached). The order parameter field is coupled with the volume fraction fields and both interfaces advance with the same speed.

The model shows some artefacts however: during evaporation, the solvent concentration in the gas phase is far from zero; this concentration strongly depends on the outgoing flux intensity. This is because the density variation of the solvent upon phase change is not taken into account, and because the diffusion coefficients used in the vapor phase are set equal to the ones in the liquid phase in order to keep reasonable simulation times. It should also be noted that some air enters into the film. This cannot be avoided with the Flory-Huggins form of the free energy: in the multicomponent equilibrium the system relaxes to, no volume fraction can be equal to zero. Nevertheless, the volume fraction of air entering the film can be limited to negligible amounts by playing with the thermodynamic parameters for air, which are actually adjustable parameters: the stronger the driving force for evaporation $F_k$ and the liquid-vapor interaction parameter for air, the less air enters into the film. However, as stated above, we expect the vapor composition and the presence of residual air in the liquid phase only to have a negligible impact on the structure formation in the liquid film.

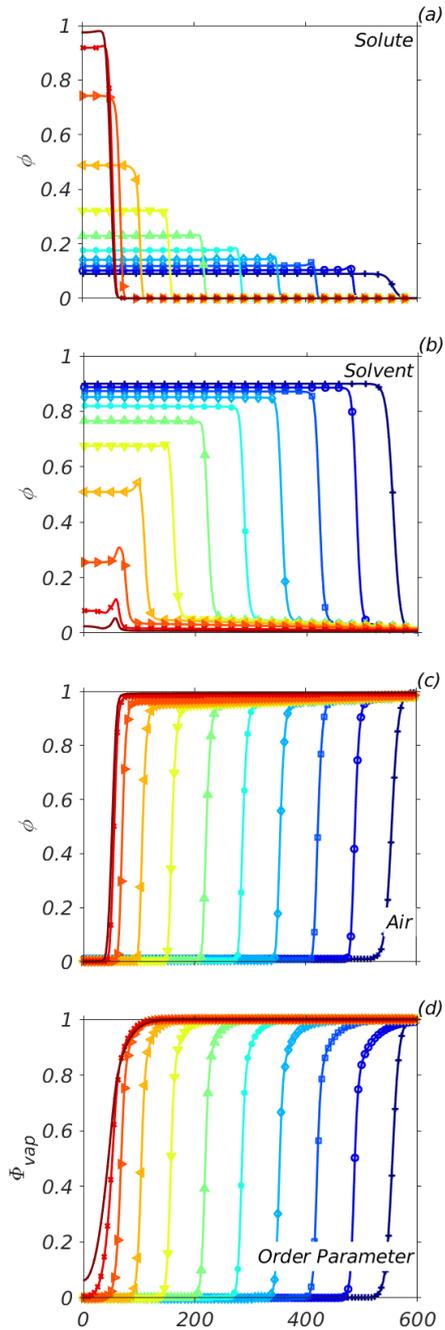

Figure 1: typical volume fraction fields for solute (a), solvent (b), air (c), and vapor order parameter (d) during evaporation. The interface moves from right to left. The time changes with time steps dt=7·10$^{-5}$s from t$_0$=0 to t$_{dry}$=7·10$^{-4}$s.

**Evaporation kinetics for low evaporation rates for the classical model and comparison with experimental data**

Before showing results of our model, we first discuss the behavior of the widely used procedure consisting of only simulating the liquid film and performing solvent evaporation through a boundary condition at the film surface. The solvent flux at the boundary is often written as

$$j_k = K(\varphi_{top,k} - \varphi_{vap,k}) \approx K\varphi_{top,k}, \quad (16)$$

with $K$ being a mass-transfer coefficient. The behavior of such a model can be solved analytically in the case of very low Biot numbers for miscible systems. Here, the Biot number characterizes the ratio between the evaporation rate and the diffusion rate and is defined as

$$Bi = \frac{h_0 K}{\Lambda}, \quad (17)$$

where $h_0$ is the initial film height and $\Lambda$ the mobility coefficient of the system. In the general case, the definition of $\Lambda$ is not straightforward because the Onsager mobility coefficients are composition dependent (see Eq. (10)-(11)) and since the materials might have different diffusion coefficients. For the sake of simplicity, we restrict the discussion to the case where all diffusion coefficients are equal and assume that $\Lambda \sim D$. The case of low Biot numbers corresponds to very low evaporation rates for which diffusion is fast enough to rule out all concentration gradients. Thus, the concentration in the film is constant and we have the zero-dimensional problem

$$\begin{cases} \frac{dh}{dt} = K\varphi_s \\ \varphi_s(t) = 1 - \frac{h_0 \varphi_{s,0}}{h(t)} \end{cases} \quad (18)$$

Here, $\varphi_s$ is the solvent volume fraction and $\varphi_{s,0}$ the initial solvent volume fraction. The second equation expresses the trivial relationship between the film height and the solvent volume fraction. Solving this system of equations, we end up with the following equation for the relative height $h_{rel}=h/h_0$:

$$h_{rel}(t) - 1 + \varphi_0 \ln\left(\frac{h_{rel}(t) - \varphi_0}{1 - \varphi_0}\right) = -\frac{K}{h_0}t = -\frac{t}{\tau} \quad (19)$$

$\tau$ is the time necessary to fully evaporate a film of height $h_0$ with the evaporation rate $K$. The results are shown in Figure 2 for various initial solvent volume fractions. The evaporation rate decreases proportional to the solvent volume fraction so that the height does not decrease linearly with time. Another noticeable effect is that for films with the same initial height and the same evaporation rate, a longer drying time is predicted for lower initial solvent quantities.

We performed drying experiments in order to check the validity of this model. We chose a polystyrene-toluene model system to ensure the miscibility of both fluids. In order to investigate the low-Biot number regime, we used a low-molecular weight polymer (Mw=35.000) from Sigma-Aldrich to have sufficiently high diffusion coefficients and performed experiments at room temperature to obtain relatively low evaporation rates. The final thickness was chosen to be in the micrometer range so that the film height can easily be measured with white light interferometry during the whole drying process. Under these conditions, assuming that, within the fast mode theory, the fast component drives the diffusion process, and that the self-diffusion coefficient of toluene in polystyrene is about $10^{-11}$ m$^2$/s, the Biot number is estimated to be around 0.1. We used initial solvent volume fractions of 60%, 70%, 80% and 90%. The wet films have been deposited by doctor blading with an applicator height of 100 µm and the applicator speed was varied to obtain an initial thickness of about 10 µm. The film thickness was monitored in-situ with a white-light interferometer Filmetrics F20-UV from Hamatsu. Refractive indexes of 1.58684 for polystyrene and 1.497 for toluene and wavelength of $\lambda_1 = 600nm$ and $\lambda_2 = 800nm$ were used to calculate the film height through $d = \lambda_1\lambda_2/2(\lambda_1 n_2 - \lambda_1 n_2)$.

The experiments have been repeated five times and normalized with the initial height and evaporation rate. The results are shown in Figure 2 in comparison with the analytical model presented above. In the experiments, the evaporation rate is almost constant up to the very last moments of the drying and the scaled evaporation times are shorter for films with a lower initial solvent concentration. The model strongly deviates from this behavior and cannot render the measured kinetics.

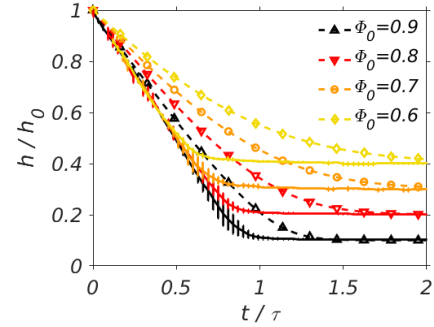

Figure 2: comparison of experimental data for a polystyrene-toluene film drying with the evaporation model (Eq. (19)) for various initial solvent volume fractions. The full lines are the experimental data and the dashed lines the model results. Error bars represent +/- 1 standard deviation

**Evaporation kinetics for low evaporation rates for the proposed model and comparison with experimental data**

In order to investigate the drying kinetics of our phase-field model in the low-Biot regime, where the concentration fields are homogeneous in the liquid film, we first investigate the impact of the phase change parameters, the driving force $F_k$ and the energy barrier $H$ on the drying kinetics (see Supplementary Information).

The normalized drying kinetics corresponding to two of the parameter sets is plotted in Figure 3. For the first set (without energy barrier and low driving force for evaporation), the drying kinetics is far from the experimental results and the evaporation rate is almost proportional to the solvent volume fraction (Figure 3a). The results are very close to the curves obtained with the analytical model presented above. Using a higher energy barrier and higher driving force, the evaporation rate is much more constant and the drying times are almost the same for all volume fractions. In general, for a more constant evaporation rate, the simulated kinetics corresponds more closely to the experimental results. This is illustrated by comparing the model results and the experimental data in Figure 3b. These results show that the proposed model has the ability to match the experimental curves almost perfectly, much better than the simple and frequently used model, (Eq. *(16)* to *(19)*). This is very important for the simulation of microstructure evolution since the drying kinetics fixes the time available for the system evolution. Moreover, it seems to indicate that an energy barrier might be involved in the evaporation process, but the physical meaning of this barrier in this simplified phase-field model is unclear at this stage. We assume that it might reflect the thermodynamics of the liquid-vapor transition and that it might be related to the heat of vaporization. Another difficulty is that the evaporation rate depends on the product of the driving force $F$ by the Allen-Cahn mobility $M$, so that $F$ cannot be determined in a unique way from the Hertz-Knudsen formula or from experiments. Both issues still require further investigation which is beyond the scope of the current work.

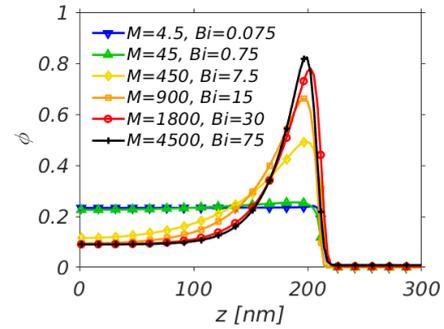

Figure 4: Solute volume fraction profile during drying at fixed height for various Biot numbers by varying the Allen-Cahn mobility coefficient.

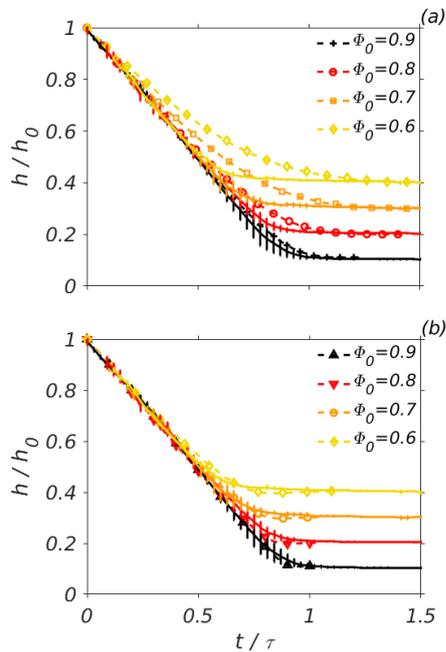

Figure 3: Simulated Evaporation kinetics (dashed lines) for different volume fractions, and compared to the experimental results (full lines). Error bars represent +/- 1 standard deviation. (a) F = 90 kJ/kg, H = 20 kJ/kg (b) F = 115 kJ/kg, H = 700 kJ/kg

**Evaporation kinetics for high evaporation rates**

At higher Biot numbers, the diffusion process is not sufficiently fast to compensate for the solvent evaporation and concentration gradients are expected to appear in the liquid film. The surface of the film is solvent-depleted because of the solvent escaping the liquid phase and a solute peak appears at the surface. The higher the evaporation rate gets, the more pronounced is the concentration peak. This is illustrated in Figure 4 where the volume fraction profiles of the solute are shown at a given film height (thus a given mean composition) for increasing Allen-Cahn mobilities, and thus increasing drying rates.

This is known to have a strong impact on the drying kinetics: with increasing Biot numbers, the evaporation is off course faster. However, the solvent surface concentration is lowered as compared to the homogeneous situation, and thus the normalized drying rate (Figure 5).

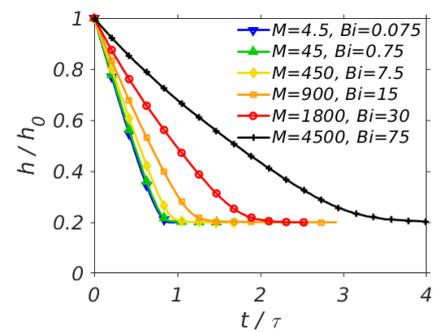

Figure 5: Simulated evaporation kinetics at high Biot numbers, for 80% initial solvent volume fraction.

**Influence of model parameters on the drying rate**

We investigate the impact of the model parameters on the evaporation rate by calculating the liquid-vapor interface velocity (see Supplementary Information). The interface velocity $V_{int,0}$ is evaluated at the beginning of the drying for very high solvent volume fraction (90%), because the evaporation rate might vary during drying as has been shown above. As expected and desired, the interface velocities vary linearly with the driving force of the solvent evaporation $F_{solv}$, and hence with the vapor pressure. It is almost proportional to the driving force when there is no energy barrier (low *H*), and the interface velocity decreases when the energy barrier increases. Moreover, the interface velocity is proportional to the Allen-Cahn mobility coefficient as expected for this kind of model. We also investigate the influence of adjustable model parameters, namely the intensity of the solvent flux boundary condition and of the "air evaporation driving force" (see Supplementary Information). The velocity of the interface decreases weakly with the solvent flux. The effect of the "air evaporation driving force" is important until the drying rate reaches an asymptotic value for sufficiently high driving forces. Under these conditions, the results obtained with the model are insensitive to this parameter.

**Two-solvent system**

Since solvent blends are often used for solution-processed photovoltaics, we also present here the ability of the model to simulate such systems. We simulate the evaporation of a film dissolved in a two-solvent mixture. The solvents only differ through the driving force for evaporation. $F_{solv1}$ is set to -90kJ/kg for the first solvent and $F_{solv2}$ is set to -55kJ/kg for the second one. Otherwise, we use the parameters reported in Table 1. From the simulation performed on single solvents, the drying rates of both solvents $V_{int1}$ and $V_{int2}$ are known to be constant and differ by a factor of 2.6. The typical volume fraction fields are shown in Figure 6. While the solute and air volume fractions evolve very similar to the situation of a single solvent system (see Figure 1), the solvent fields show interesting features: while the volume fraction of the fastest solvent decreases, the volume fraction of the slowest solvent in the film first increases and then decreases (see also Supplementary Information). This volume fraction increase is only seen if the evaporation rate difference between both solvents is sufficiently large.

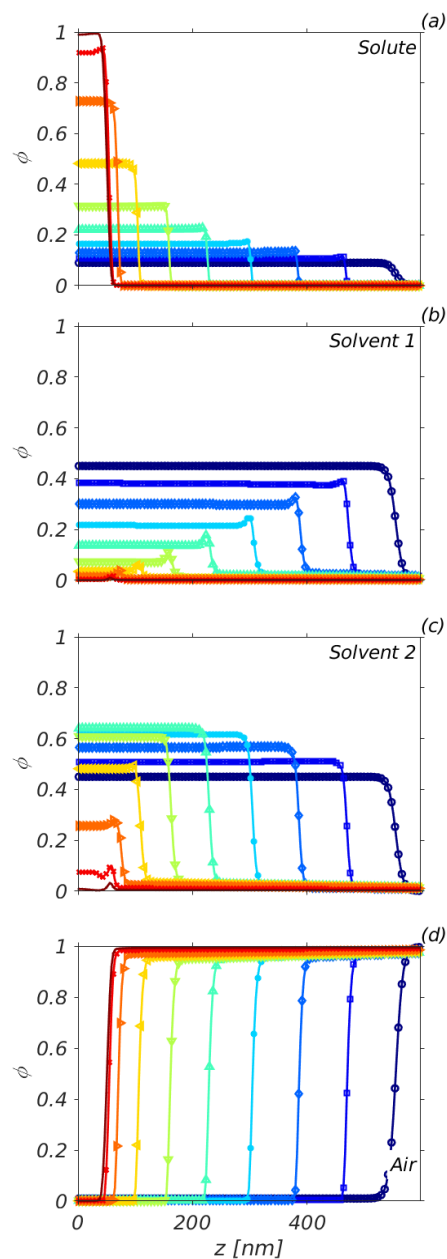

Figure 6: Volume fraction fields for a two-solvent system during evaporation. (a) Solute (b) 1st solvent (c) 2nd solvent (d) Air. The interface moves from right to left. The time changes with time steps $dt=10^{-4}$s from $t_0=0$ to $t_{dry}=10^{-3}$s. $F_{solv1}$ = -90kJ/kg and $F_{solv2}$ = -55kJ/kg

Moreover, in this particular case where both solvents have the same surface tension, there is a maximum of the volume fraction at the film surface for the fastest evaporating solvent. The drying kinetics is also remarkable. Although the evaporation rate of both solvents individually is almost constant during the whole drying, the global evaporation rate of the solvent blend is far from constant (see Figure 7 and Supplementary Information). This is because the mixing term of the free energy is more important in a three-component system than in a two-component system, and therefore the ratio of the driving force for evaporation to the driving force for mixing is smaller. This is then similar to the situation of a smaller driving force for evaporation in a single solvent system, which has been shown to favor evaporation rates proportional to the solvent volume fraction.

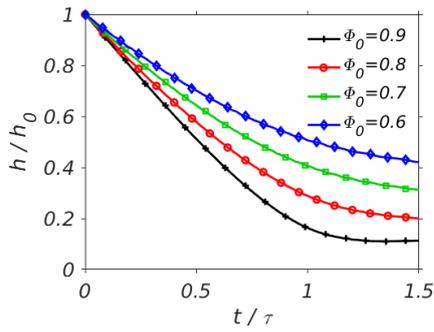

Figure 7: Normalized drying kinetics of the two-solvent system for different initial volume fractions

**Solvent vapor annealing**

Solvent vapor annealing or swelling of a dry structured film are processing approaches often used in the field of polymer solutions or solution processed electronics. This can help improving the film structure after the first drying step. Here, we demonstrate the ability of the model to simulate such a physical process. In Figure 8, we show the volume fraction profiles obtained by simulating such a swelling process, whereby the solvent partial pressure in the environment is supposed to be higher than the vapor pressure and kept constant. This means that the driving force $F_{solv}$ is now positive. Compared to the parameters presented in Table 1, only the driving force (changed to $F_{solv}$ = +55kJ/kg) and the solvent partial pressure ($\varphi_{vap} = 0.1$) are changed, and the simulation is started from a quasi-dry film. It can be seen that the film now grows, with the solute volume fraction decreasing and the solvent volume fraction increasing. Almost no air is present in the liquid film. Note that here again the fields in the vapor phase might be quite different from reality. In comparison to a drying simulation (Figure 1), the solute concentration peak is at the surface is more pronounced.

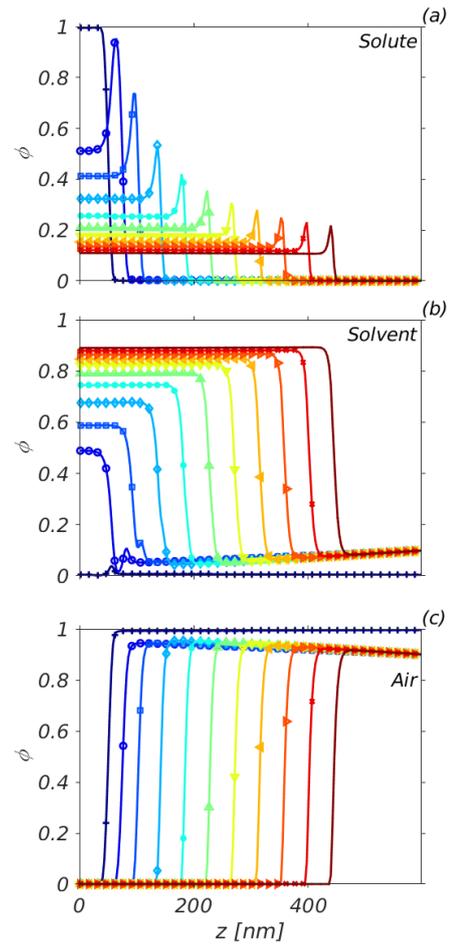

Figure 8: Volume fraction fields for (a) solute (b) solvent (c) air during a swelling simulation. The interface moves from left to right. The time changes with time steps dt=3·10$^{-5}$s from $t_0$=0 to $t_{dry}$=3·10$^{-4}$s.

## Simulation of a rough film

In this section, we illustrate the ability of the model to simulate the potential roughness of the final dry film. The first example deals with the drying of a film on a rough substrate (Figure 9). Once again, the parameters reported in Table 1 are used for a 2D simulation of a wet film drying. However, here vertical structures are present on the substrate. The grid size is here 300*100 mesh points and the grid resolution 1nm. The position of the liquid-vapor interface is shown at different times in Figure 9. The interface bends around and wraps the structures until the film is dry. The interface curvature is related to the capillary forces due to the surface tension.

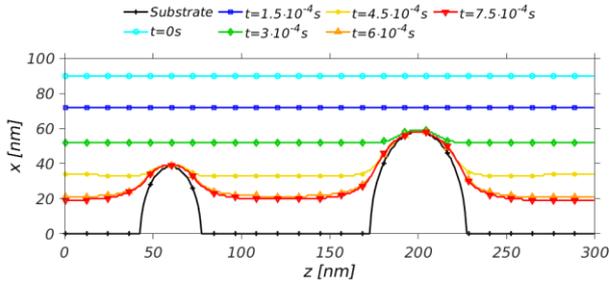

Figure 9: position of the interface of a liquid film drying on a rough substrate at different stages of the drying.

The second showcase deals with evaporation-induced spinodal decomposition (Figure 10 to Figure 12) Instead of a single solute, there are two immiscible solutes in the system. The parameters for this simulation are given in Table 2, except for the Flory-Huggins interaction parameter between the two solutes which is set to $\chi_{12,ll} = 5$. With this value, the system is expected to demix as soon as the solvent volume fraction goes below 60%. The simulation is started with a slightly higher volume fraction and with a Gaussian random noise added to the Cahn-Hilliard equation. The grid size is 264*200 mesh points and the grid resolution 1.5nm.

| T | 300 K | $F_{solv}$ | -90 kJ/kg |
|---|---|---|---|
| $v_0$ | $2 \cdot 10^{-4}$ m³/mol | $F_{air}$ | -300 kJ/kg |
| $N_i$ (all fluids) | 1 | $H_{solv}$ | 500 kJ/kg |
| $\rho_i$ (all fluids) | 1000 kg/m³ | $H_{air}$ | 500 kJ/kg |
| $\chi_{ij,ll}$ (all but $\chi_{12,ll}$) | 0 | $\varepsilon_{vap}$ | $1.3 \cdot 10^{-5}$ (J/m)$^{0.5}$ |
| $\chi_{kj,vl}$ (all) | 5 | M | $10^6$ s$^{-1}$ |
| $\chi_{kj,vv}$ (all) | -10 | $\varphi_{vap}$ | 0 |
| $\kappa_i$ | $\kappa_{i=1...3} = 2.5 \cdot 10^{-10}$, $\kappa_4 = 5 \cdot 10^{-10}$ J/m | b | 20 J·m$^{-4}$ |
| $D_{s,i}^{\varphi_k \to 1}$ (all) | $2 \cdot 10^{-9}$ m²/s | | |

Table 2: parameter set used in the 2D simulations of spinodal decomposition

We obtain the phase-separated film in Figure 10 when all the solvent has evaporated. It can be clearly seen that the film is rough (with a 30nm-high structure) although the system is completely symmetric, without any difference between both solute's molar volumes, interaction properties with the solvent or diffusion coefficients. This is due to the fact that the Biot number resulting from the chosen parameters is relatively high (*Bi=1.35*), so that the diffusion process is not fast enough to equilibrate all concentration gradients resulting from the evaporation. In particular, in the immiscible film, separated phases appear randomly and may lead to irregular patterns (see Figure 10). Surface tension effects around these irregularities generate the roughness of the surface, for which diffusion has no time to compensate. As a result, the dry film is still rough.

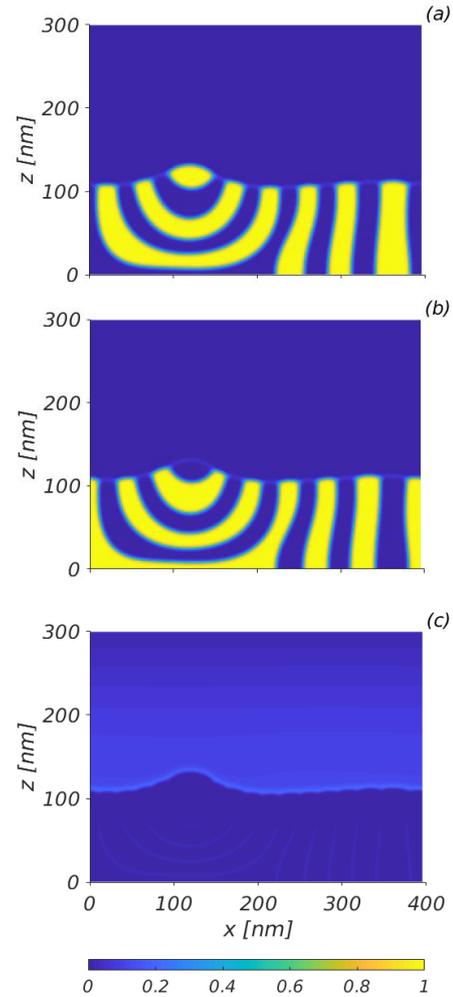

Figure 10: 2D simulation of evaporation-induced spinodal decomposition: volume fraction fields of the almost dry film. (a) 1$^{st}$ solute, (b) 2$^{nd}$ solute, (c) solvent

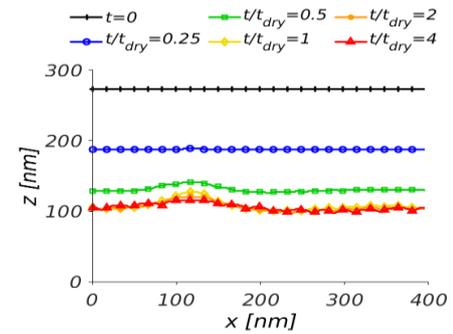

Figure 11: 2D simulation of evaporation-induced spinodal decomposition: position of the film surface with increasing time. $t_{dry}$ is the time from which the volume fraction of solvent in the film is less than 2%

This is however a transient state. With increasing time and further diffusion, the system minimizes its global energy and hence the interface density in the film, so that the roughness decreases after the drying. This is illustrated in Figure 11, showing the surface profile of the film during and after evaporation of the solvent: the height of the structure has decreased by 50% after a time equivalent to four times the drying time and will decrease further.

Figure 12 shows the arithmetic average surface roughness $R_a$ of the film during and after evaporation of the solvent for different parameter sets. The black curve corresponds to the simulation shown in Figure 10. If the evaporation is slower or the interaction parameter lower (green and blues curves), the surface tension of the film has time to compensate for any roughness pattern that could be generated through the spinodal decomposition and the roughness remains negligible. Furthermore, diffusion coefficients usually depend on the mixture composition: they decrease with decreasing solvent volume fraction, which might result in a quenching of the roughness of the film. To illustrate this effect, we simulated a system where the diffusion coefficients decrease proportionally to the solvent volume fraction from $2 \cdot 10^{-9}$ m$^2$/s in the pure solvents to $2 \cdot 10^{-10}$ m$^2$/s in the pure solutes (red curve). The generated roughness during the drying is much stronger than in the case with constant diffusion coefficients and drops quickly at the end and after the total drying, because the initially higher roughness generates higher surface forces.

Note that on these test cases, the roughness decreases in a relatively short time range because the diffusion coefficients of the solutes are very high, namely identical to the one of the solvent. For much more viscous materials, the generated roughness would need much longer times to decrease. Furthermore, in polymer systems, the diffusion coefficients actually decrease by several orders of magnitude with decreasing solvent concentration and might be very low in the dry film. Thus, if the drying has been fast enough to generate significant height variations, these height variations might be kinetically quenched for a long time. If the polymeric film is kept away from the viscous regime, for instance if the solutes are kept below their glass transition temperature, these height variations might even remain stable. This is one possible sources of roughness observed in solution-processed polymer films.

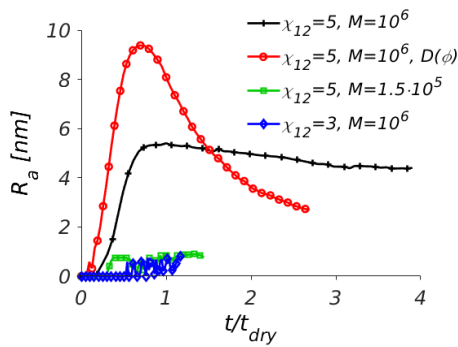

Figure 12: Effect of the interaction parameter, of the Allen-Cahn mobility and diffusion coefficient concentration dependence on the arithmetic average surface roughness $R_a$

## Conclusion and perspectives

In this paper, we have presented a new evaporation model for the modelling of drying films within the phase-field simulation framework. This allows to describe how an evaporating film evolves kinetically towards their thermodynamic equilibrium. Our goal is to propose a model that would be able to handle the film roughness, to match the experimental drying curves, and that would be relatively simple at the same time. These are important requirements to properly represent and focus on the structure formation processes such as liquid-liquid demixing and crystallization in complex mixtures.

The developed simulation code is three-dimensional, although only 1D and 2D simulations have been shown in the present contribution, and it can handle any number of solvents or solutes. In contrast to a number of existing simulation tools, the vapor phase is explicitly present in the system which enables to investigate roughness effects. Strong simplifying assumptions have been made in order to keep the model simple; in particular, the density variation upon phase change and hydrodynamics are not taken into account. This implies notably that the description of the vapor phase is not correct, which is affordable because our target is only the description of the liquid film, which is not affected by these assumptions.

The evaporation model is based on a classical phase-field description of phase transition, whereby the driving force for the liquid-vapor interface evolution is proportional to the difference between the solvent vapor pressure and partial pressure in the environment, following the original Hertz-Knudsen formula. Among others, this ensures a correct dependence of the drying rate to the solvent vapor pressure. An energy barrier is also present for the liquid-vapor transition, and it has been shown that the drying kinetics strongly vary with the value of the driving force and the energy barrier: the higher the driving force and the energy barrier, the more constant is the evaporation rate. This is of primary importance, since the evaporation rate has been experimentally shown in this paper to be constant during the whole drying of a polystyrene-toluene system. Hence, our model has the ability to match very nicely experimental drying curves, in contrast to models basing on the classical assumption that the drying rate is proportional to the solvent volume fraction at the film surface. The ability of the model to simulate also film swelling has been demonstrated.

On the one hand, future work will be devoted to further investigations of the model's behavior and to comparison with experiments. The sensitivity of the drying curves to the driving force of the evaporation (or solvent vapor pressure) and to the heat of vaporization could be measured in order to check the validity of the modelling results presented in this work. This would also help understanding the physical meaning of the energy barrier proposed in our model and relate it to solvent measurable properties. The experimental investigation of drying curves for solvent blends will also be a good test for the model. Additionally, the influence of other model parameters such as molar volumes, surface tensions, fluid interactions and diffusion coefficients will be further investigated in order to gain full understanding of the model before going towards the simulation of real systems.

The presented phase-field framework will be used for applications in the fields of solution-processed photovoltaics. Organic photovoltaic layers will be investigated first. In these wet-deposited films, two immiscible active materials undergo a liquid-liquid phase separation and form a nanometer-scaled

"bulk heterojunction" structure which is necessary for the photovoltaic performance. We will investigate the impact of the processing conditions and solvent properties on the final film structure. There, we will deal first with amorphous material systems, with a focus on the impact and role of solvent blends, which have barely been theoretically studied until now. This can be readily done with the framework presented here, as it has been shown on a very simple case. Second, the formation of perovskite polycrystalline layers will be investigated. The challenge is to include the crystal nucleation and growth processes in the model, but this could be done with minor changes in principle. This will be the topic of a future paper. Finally, this will allow us to also study organic semi-crystalline layers, which make up most of the high performing organic solar cells.

## Conflicts of interest

There are no conflicts to declare.

## Acknowledgements

The authors acknowledge financial support by the Deutsche Forschungsgemeinschaft (DFG) within the Cluster of Excellence "Engineering of Advanced Materials" (project EXC 315) (Bridge Funding).

## Notes and references